\documentclass[printer]{aa}
\usepackage{graphicx}
\usepackage{txfonts}

\usepackage{tipa}
\usepackage{amsmath}
\usepackage{mathrsfs}
\usepackage{natbib,twoopt}
\usepackage[breaklinks=true]{hyperref} 
\bibpunct{(}{)}{;}{a}{}{,} 
\pdfoutput=1

\begin{document} 

\def \pg {PG1211+143}
\def \Sw {{\it Swift}}

\title{Modelling the variable broad-band optical/UV/X--ray spectrum of \pg: Implications for the ionized outflow}
      \author{I. E. Papadakis\inst{1,2}
          \and
          F. Nicastro\inst{1,3,4}
          \and
          C. Panagiotou\inst{1}
          }
   \institute{Department of Physics \& Institute of Theoretical \& Computational Physics, University of Crete, PO Box 2208, GR-710 03 Heraklion, Crete, Greece 
         \and IESL, Foundation for Research and Technology-Hellas, GR-71110 Heraklion, Crete, Greece 
         \and Osservatorio Astronomico di Roma-INAF, Via di Frascati 33, I-00040 Monte Porzio Catone, RM, Italy   
         \and Harvard-Smithsonian Center for Astrophysics, Cambridge, MA 02138, USA}

\authorrunning{I. Papadakis et al.}
\titlerunning{Opt/UV/X--ray variability of \pg}
\date{Received .. ...... ; accepted .. ...... }

\abstract
 {We present the results from a detailed analysis of the 2007 \Sw\ monitoring campaign of the quasar \pg.}
   {We study its broad-band optical/UV-X--ray spectral energy distribution and its variations, with the use of physically motivated models.}
   {We constructed broad-band, optical/UV-X--ray spectral energy distributions over three X--ray flux intervals, and we fitted them with a model which accounts for the disc  and the X--ray coronal emission. We also added a spectral model component to account for the presence of the warm absorber which has been well established from past observations of the source. }
   {We detected no optical/UV variations over the two-month period of the monitoring campaign. On the other hand, the X--rays are highly variable in a correlated way in the soft and hard X--ray bands with an amplitude larger than has been commonly observed in nearby Seyferts, even on longer time scales.  The three flux spectra are well fitted by the model we considered. The disc inner temperature remains constant at $\sim 2$ eV, while X--rays are variable in slope and normalization. The absorber covers almost 90\% of the central source. It is outflowing with a velocity less than 2.3$\times10^4$ km/s (3$\sigma$ upper limit), and has a column density of log N$_H \sim 23.2$. Its ionization parameter varies by a factor of 1.6, and it is in photo-ionizing equilibrium with the ionizing flux. It is located at a distance of less than 0.35 pc from the central source, and its relative thickness, $\Delta R/R$, is less than 0.1. The absorber's ionization parameter variations can explain the larger than average amplitude of the X-ray variations. }
 {The absence of optical/UV variations are consistent with the high black hole mass estimate of $\sim 10^8$ M$_{\odot}$ for this object, which implies variability time scales longer than the period of the \Sw\ observations. It argues against the presence of inward propagating fluctuations in the disc as the reason for the flux variability in this source and against the hypothesis that X--ray illumination singificantly affects the disc emission. This is consistent with the low ratio of X--ray over the bolometric luminosity of $\sim 20-35$ in this source. Based on the properties of the ionized outflow, we estimate an upper limit for the mass outflow of $\sim 5$ M$_{\odot}$ per year, which is $\sim 2.3$ times the Eddington mass  accretion rate for \pg. If the outflow rate is indeed that high, then it must be a short-lived episode in the quasar's life time. Finally, we estimate an upper limit for the kinetic power of the outflow of  $\sim 1.4\times 10^{43}$ ergs s$^{-1}$. As a result, the outflow cannot deploy significant mechanical energy to the surrounding ISM of the quasar's host galaxy, but is sufficient to heat the ISM  to $10^7$ K and to produce a fast decline to the star formation rate of the galaxy.   }

\maketitle
%

\section{Introduction}

It is generally accepted that active galactic nuclei (AGN) are powered by the accretion of matter onto  supermassive black holes (BHs) situated at the centres  of galaxies in the form of a geometrically thin, optically thick disc. Assuming black-body  emission of the locally dissipated energy in the disc,  it is expected that the optical/UV (opt/UV) spectra of AGN are dominated  by a broad quasi-thermal feature with an inner disc radius temperature of the order of  $ \sim 10^4-10^5$K (e.g. Shields, 1978). This feature has been detected in the opt/UV  part of the broad-band energy distributions of bright quasars (e.g. Elvis et al. 1994), and is commonly referred to as the AGN ``big blue bump''. In addition to this emission component, X--ray emission is also a ubiquitous feature of the AGN spectra. Variability arguments and microlensing observations (e.g. Mosquera et al. 2013) indicate that the X--ray source size is  less than $\sim 10-20 r_{\rm g}$ (where $r_{\rm g}$ is the gravitational radius), and is probably located very close to the central BH where most of the gravitational energy is liberated. The main process for the production of the X-rays is generally thought to be the Comptonization of soft photons (like the ones emitted by the disc) by electrons of a hot (kT$_e \sim 100 - 150$ keV) corona. 

If the above picture is correct, the opt/UV and X--ray emission should be somehow related in AGN. For example, if the seed photons for Comptonization are those emitted by the disc, their variations should affect the cooling/heating of the X-ray corona, and hence result in X-ray spectral variations. Furthermore, given that the AGN emitted power is provided by matter accretion, one expects variations of the accretion matter rate to affect both the opt/UV and the X--ray region. If the X--ray source is located in the innermost region in these objects,  the variations are expected to originate at longer wavelengths (optical) and then move to shorter ones (UV, X-rays). Finally, a fraction of the X-ray emission may be intercepted and reprocessed by the material in the disc, serving as an external heating source for the disc. In this case,  the X--ray variations are expected to preceed those observed in the opt/UV band with a delay proportional to the light travel time from the X--ray source to the disc region responsible for the opt/UV emission.

An additional physical component, namely absorbing material which lies along the line of sight towards the central engine of  AGN, is revealed by the detection of absorption lines in the UV and soft X--ray spectra of AGN. Approximately 50\% of Seyferts and quasars show absorption signatures in the UV band (e.g. Crenshaw et al. 1999) with a similar fraction of AGN showing warm absorbing material signatures in their X-ray spectra (e.g. George et al. 1998; Piconcelli et al. 2005). Both the UV and the soft X--ray band absorbers appear to outflow. The warm absorber velocities are rather low, of the order of a few hundred km/s, and their kinetic power is less than 1\% of the AGN bolometric luminosities (Blustin et al. 2005; Krongold et al. 2007). However, over the last few years highly ionized, high-velocity outflows (with a velocity greater than $10^{4}$ km/s) have been detected in a few AGN. They are manifested as blue-shifted K-shell absorption lines from Fe XXV and Fe XXVI at energies E$>7$ keV (e.g. Tombesi et al. 2013, and references therein). These ultra-fast outflows may contribute significantly to the AGN feedback to their host galaxy.

We present the results from the analysis of the \Sw\  monitoring data of the quasar \pg. This is a Type I AGN at a redshift of $z=0.0809$ (Marziani et al. 1996). Its optical emission lines are narrow;  the FWHM of the H$_{\rm \beta}$ line is 1832$\pm$81 km/s (Kaspi et al. 2000). This has led to the classification of this object as a Narrow Line Seyfert 1 (NLS1) galaxy. \pg\ shows a strong big blue bump   and a bright X-ray emission. It is one of the first AGN where an ultra-fast ionized outflow with a velocity of $\sim 0.09 c$ has been proposed (Pounds et al. 2003). Kaspi \& Behar (2006) have analysed both the CCD and the RGS data of the same 2001 {\it XMM-Newton} observation and report the presence of an outflow component of about 3000 km/s, in contrast with the ultra-fast velocity reported by Pounds et al. (2003). Further evidence of the fast ionized outflow has been reported by Pounds \& Reeves (2007, 2009) using additional XMM-Newton data. Zoghbi et al. (2015) did not detect absorption features indicative of a fast outflow  in a recent NuSTAR observation.  In a recent paper, Pounds (2014) has reported the results from  a new analysis of the XMM-Newton RGS data, which suggests the presence of two absorbers with outflow velocities of $\sim 0.07$ and $0.14$c. 

\Sw\ observed \pg\ several times over a period of just over two months, in March--May 2007. Bachev et al. (2009)  have already studied the $\Sw$ data of the source from the monitoring campaign in 2007. They found that the source was highly variable in X--rays but showed little variations in the opt/UV bands. They also found that its X--ray spectrum can be explained by either a partial covering absorber or by X-ray reflection onto the disc, but they were not able to distinguish between the two scenarios. 

We model in detail the broad-band (optical/UV/X--ray) spectral energy distribution (SED) of the source. Our main aim is to investigate whether a simple physical model, consisting of an accretion disc, a hot corona, and a warm absorber, can fit the broad-band SED of the source. The \pg\ data from the 2007 \Sw\ monitoring are particularly useful for this purpose as the X--ray flux was highly variable during the campaign, while the optical/UV flux remained constant  at all flux intervals. This observational fact can put strong constraints on the modelling of the broad-band SED, simultaneously, in all the different X--ray flux intervals of the source. 

\section{Observations and data analysis}
%

\begin{table}
\caption{\Sw\ observations log.}         
\label{table:1}    
\centering           
\begin{tabular}{lccccc}      
\hline\hline               
        & $T_{\rm start}$        &                   &       $T_{\rm exp}$(s)            &    &         \\  
        &  (Date/UT)    &(V,B,U)        & (UVW1)        & (UVW2)        & XRT \\  \hline                      
1       & 09-03/01:00   & 66.7  & 133.6         &  268.5        &  1766 \\
2       & 10-03/17:18   & 37.2  & 74.6  &150.4  & 1688 \\
3       & 11-03/19:01   & 37.2  & 75.6  & 150.3 & 1691 \\
4       & 12-03/15:54   & 37.2  & 74.6  & 150.4 & 1656 \\
5       & 13-03/20:49   & 37.2  & 74.6  & 150.4 & 829.1 \\
6       & 14-03/09:07   & 55.9  & 112.0 & 223.2 & 1973 \\
7       & 15-03/11:05   & 81.5  & 163.2 & 325.6 & 1748 \\
8       & 16-03/00:00   & 27.3  & 55.9  & 111.0    & 1930 \\
9       & 17-03/08:20   & 27.3  & 55.9  & 111.0    & 2677 \\
10      & 18-03/10:04   & 27.3  & 55.9  & 111.0    & 1721 \\
11      & 19-03/10:10   & 22.4  & 46.0  & 91.3  & 1671 \\
12      & 26-03/07:06   & 101.2 & 202.5 & 405.3 & 2000 \\
13      & 02-04/20:56   & 114.0 & 228.1 & 456.5 & 2442 \\
14      & 11-04/09:11   & 39.1  & 79.5  & 158.2 & 1741 \\
15      & 17-04/00:11   & 45.0  & 90.3  & 181.9 & 2030 \\
16      & 22-04/11:52   & 56.9  & 113.9 & 228.1 & 2055 \\
17      & 30-04/04:43   & 56.9  & 113.0 & 227.1 & 1873 \\
18      & 09-05/15:12  & 69.7   & 138.5 & 261.9 & 1393 \\
19      & 14-05/04:02  & 90.3   & 180.9 & 361.9 & 2255 \\
20      & 20-05/01:27  & 83.4   & 167.1 & 64.7  & 839.1 \\
\hline                                 
\end{tabular}
\end{table}

PG1211+143 was observed by \Sw\ twenty times from 9  March 2007 to 20  May 2007. At first, \Sw\ observed the source daily; after March 19, Swift observed \pg\ every 5 to 9 days. A summary of these observations is given in Table 1. Columns 1 and 2 list the number, the date, month, and the start time (UT) of each observation. The next four columns list the exposure time of each UVOT filter and XRT in seconds. We note that the source was not observed in the $V-$band during the last observation (20 May  2007).

The \Sw\ XRT (Burrows et al. 2005) observations were performed in Photon Counting mode. The event file of the observation was created by using the \Sw\ analysis tool {\tt xrtpipeline}. Source photons were extracted from a circle with a radius of 20 pixels centred on the source. The background was selected from a nearby, source-free region with a radius of 100 pixels. Only single to quadruple events in the energy range of 0.3-10 keV were selected for further analysis. Source and background spectra were extracted from the event file by using XSELECT version 2.4b. The auxiliary response files were created by the Swift tool {\tt xrtmkarf}. We used the response matrix version 012 with a grade selection 0-12.

In addition to the X-ray data, we also obtained photometry with the UV/Optical Telescope (UVOT; Roming et al. 2005) in the V, B, U, UVW1, and UVW2 filters.  \Sw\ observed the source with the UVM2 filter as well. Since its effective area overlaps  the effective area of the UVW1 and UVW2 filters, the respective source's flux will be heavily correlated with the flux measured with the UVW1 and UVW2 filters. For this reason, we do not consider the data from the UVM2 filter. Photons were extracted from a circular region with r = 5 arcsec, and the background from a circular region with a radius of 50-75 arcsec located next to the source extraction region. The UVOT tool {\tt uvotsource} was used to determine the magnitudes and fluxes (Poole et al. 2008; Breeveld et al. 2010). The fluxes were corrected for Galactic reddening (E(B-V)=0.035; Schlegel et al. 1998) with the standard reddening correction curves by Cardelli et al. (1989), as described by equation 2 in Roming et al. (2009).

\section{ Optical/UV and X--ray light curves}

\begin{figure}
\centering
\resizebox{\hsize}{!}{\includegraphics{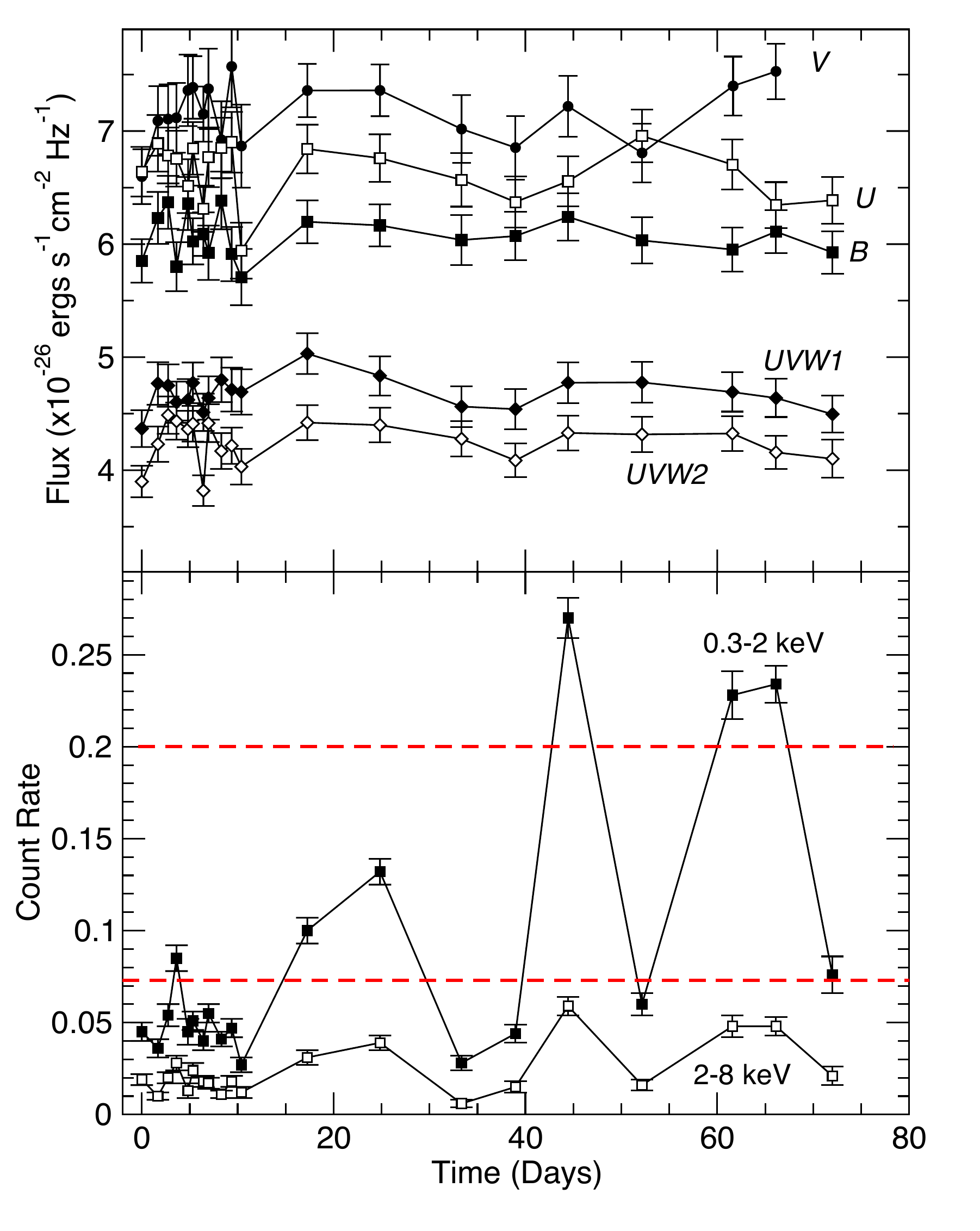}}
\caption{ \pg\ UVOT and X--ray light curves (top and bottom panels, respectively) during the 2007 \Sw\ monitoring campaign. }
\label{Fig1}
\end{figure}

The top panel in Fig.\,1 shows a plot of the \pg\ light curves in all the UVOT bands (a value of 0.5$\times 10^{-26}$ erg s$^{-1}$ cm$^{-2}$ Hz$^{-1}$ has been added to the V filter's fluxes for clarity). The bottom panel shows the X--ray light curves in the 0.3--2 and 2--8 keV band (soft and hard X--rays, respectively). Time on the {\it x-}axis is measured in days starting from the date of the first observation.

The UVOT band light curves exhibit variations which are within the observational errors. Application of the $\chi^2$-test confirms that the the V, B, U, and UVW1 light curves are consistent with the hypothesis of a constant flux (the probability, $p$, is larger than 0.22 in all cases). The probability of constant flux is somewhat lower for the UVW2 light curve ($p=0.06$), but this is still not low enough to claim significant variations. The UVW2 intrinsic fractional root mean square variability amplitude ($F_{\rm rms}$) is $0.02\pm 0.01$ (the errors correspond to the measurement errors only, and are computed as described in Vaughan et al. 2003). The X--ray light curves are highly variable. We observe the typical, erratic variations we observe in the long term light curves of other AGN as well.  The observed variations show a soft and hard band min-to-max variability amplitude of the order of  $\sim 10$ and $\sim 5.5$, respectively. We found that $F_{\rm rms}=0.86\pm 0.02$ and $0.59\pm 0.03$ in the soft and hard X--ray bands, respectively. 

Given  that we do not detect significant variations in the optical/UV bands, it is meaningless to search for X--ray/opt/UV correlations. On the other hand, the soft and hard band X--ray variations are highly correlated. This is demonstrated in Fig.\,2, where we plot the soft vs the hard band count rate.  We fitted a straight line (i.e. $y=a+bx$) to the data using the subroutine {\tt fitexy} of Press et al. (1992) in order to take into account the errors on both count rates. A straight line fits the data reasonably well:  $\chi^2=25.7/18$ degrees of freedom ($p_{\rm null}=0.11$). However, the best-fit $y-$intercept turned out to be negative ($a=-0.024\pm 0.007$), which does not have an obvious physical explanation. A straight line also provides an acceptable fit when we consider the data with a hard count rate smaller than 0.04 counts/s ($\chi^2=14.2$ for 15 degrees of freedom). In this case, the best-fit $y-$intercept is consistent with zero: $-0.015\pm0.007$ counts/s. This is the case (good fit and best-fit $y-$intercept consistent with zero) irrespective of the upper limit on the $2-8$ keV band count rate we consider when we fit the data, as long as it is smaller than 0.04 counts/s. 

The red dashed line in Fig.\, 2 indicates the best-fit line to the data with 2--8 keV count rate smaller than 0.04 counts/s. Clearly, the highest X-ray flux points are not consistent with this line. They lie above the extrapolation of the best-fit line to higher count rates. Our results suggest that the relation between the soft and hard band count rates is linear, but at higher fluxes the soft band varies with a larger amplitude than the hard band. The value of the upper count rate of 0.04 c/s may seem arbitrary; however, as we discuss in \S 5.3, this behaviour (i.e. non-linear  above a certain X--ray {\it luminosity} level) could arise in the case of changes in the opacity of a warm absorber, which would affect mainly the soft band photons.

\begin{figure}
 \resizebox{\hsize}{!}{\includegraphics{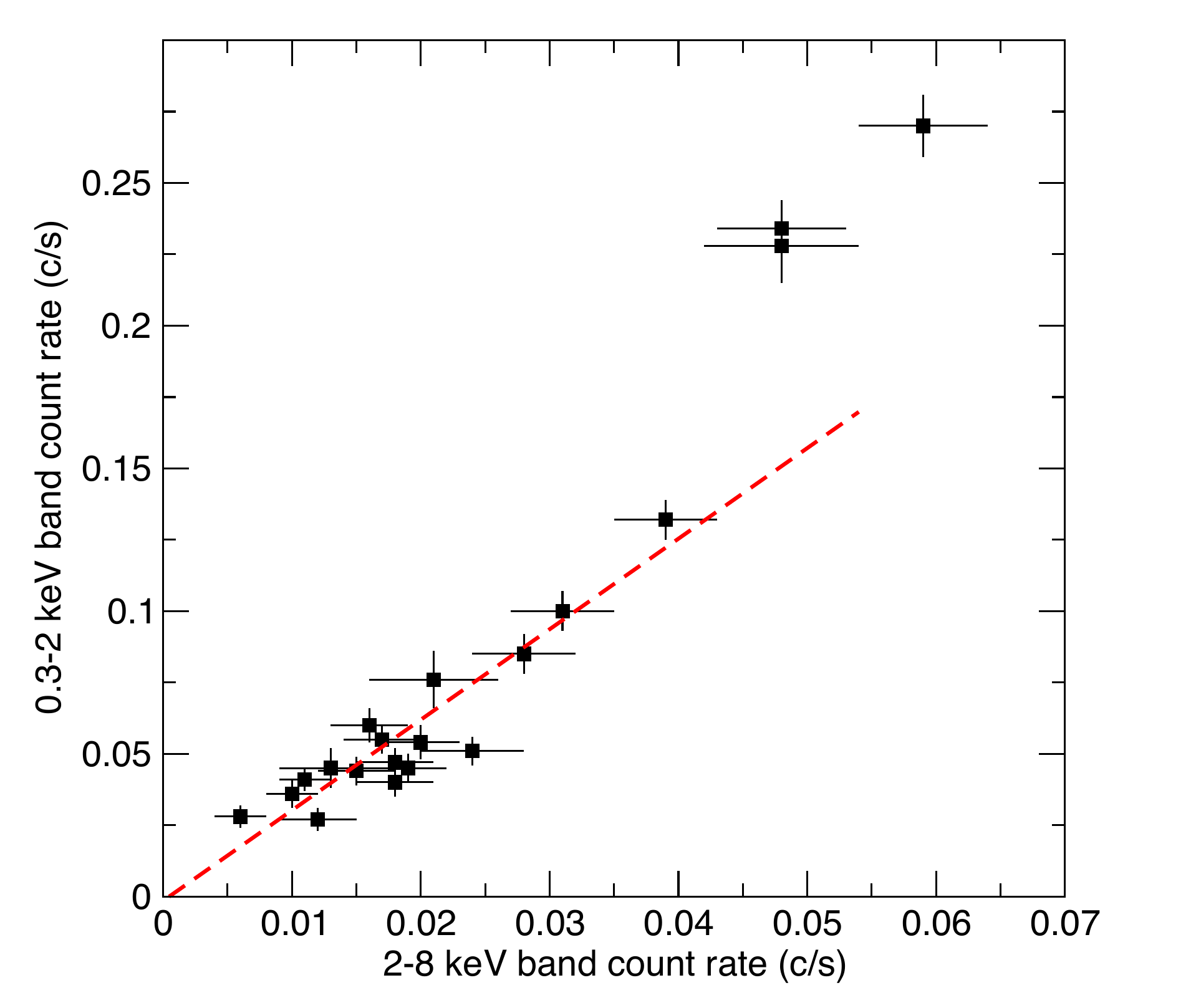}} 
\caption{ 0.3--2 keV vs the 2--8 keV band count rate. The dashed line indicates the best-fit model to the data with a hard band count rate less than 0.04 c/s.}
 \label{FigVibStab}
  \end{figure}

\section{Flux resolved spectroscopy}

In order to study the broad-band spectral energy distribution of the source (SED), we defined three different flux intervals based on the  observed 0.3-2 keV light curve: (a) a high flux-interval (HF: 3 Swift observations with a soft band count rate $> 0.2$ ct s$^{-1}$), (b) a medium flux-interval (MF: 4 Swift observations with count rates in the range 0.07-0.2 ct s$^{-1}$), and (c) a low flux-interval (LF: 14 Swift observations with count rates $<0.07$ ct s$^{-1}$). The horizontal dashed lines in the bottom panel of Fig.\,1 indicate the flux limits for these intervals. The three HF observations are those which do not follow the linear trend indicated by the dashed line in Fig.\,2. They were taken in the last 25 days of the monitoring campaign. The MF state data correspond to observations that were taken throughout the observing \Sw\ run. The LF state data include most of the observations that were mainly taken during the first intensive monitoring period of the \Sw\ campaign. We note that these intervals refer to distinct {\it X--ray} flux intervals of the source and not to its overall flux output which, given the lack of opt/UV variations,  remains almost constant throughout the \Sw\ observing run.
 
To create the total X--ray spectrum for each state, we used {\tt mathpha} to combine the individual source and background spectra of all the observations, within the respective state. The X--ray spectra were then rebinned with {\tt grppha} (v. 3.0.1) to have at least 15 photons per bin. We computed the mean flux in each UVOT band using the deredenned mean flux measurements of the individual observations within each state. We then used the {\tt flx2xsp} command to transform the flux measurements to count rates, and hence to create the respective opt/UV spectrum. For the spectral analysis we used the fitting package {\tt   Sherpa} (Freeman et al. 2001), part of the {\tt   Chandra} Interactive Analysis of Observation software (CIAO; Fruscione et al. 2006).

\subsection{Disc plus corona model}

We first modelled the broad-band SED (extending over four orders of magnitude in energy)  with a single spectral component, namely {\tt   xsnthcomp} in {\tt   Sherpa} (Zdziarski, Johnson \& Magdziarz 1996; Zycki, Done \& Smith, 1999), attenuated (only in the X-rays) by the amount of Galactic neutral absorption along the line of sight to PG~1211+143, $N_H = 2.7 \times 10^{20}$ cm$^{-2}$ (Kalberla et al. 2005). This model provides a description of  the continuum shape from thermal Comptonization. The model parameters are  (a) the electron temperature, T$_e$, of the hot Comptonizing corona, which determines the high-energy cut-off of the X-ray spectrum; (b) the photon index, $\Gamma$, of the Comptonized X--ray powerlaw; (c) the redshift, $z$, of the emitting source; (d) the model normalization, $N$; and  (e) the soft-photon temperature, $T_{BB}$. We chose the seed photons to have a multicolour, disc black-body energy distribution in order to  simultaneously fit  the opt/UV and the X--ray data with the same model.

In practice, we did this by fitting the six opt/UV and X-ray spectra (two spectra for each of the HF, MF, and LF intervals) with six distinct 
 {\tt   xsnthcomp}  model components. For each of the three flux-state spectra, all parameters of the three X-ray {\tt   xsnthcomp} components are linked to the respective parameters of the associated UV {\tt   xsnthcomp} components. For all spectra, the high-energy cut-off of the Comptonized X-ray power law (to which our data are insensitive with the current UVOT+XRT data) is frozen to T$_e = 100$ keV \footnote{Zoghbi et al. (2015) have recently reported a lower limit of 124 keV on T$_e$. Our results remain unaffected even if we choose a value as large as 200 keV for T$_e$}. The only three parameters free to vary in the fit are therefore T$_{BB}$, $\Gamma$, and $N$. Since we do not observe any significant opt/UV flux variability during the \Sw\ campaign, we force T$_{BB}$ to be the same for all  three flux-state spectra, while $N$ and $\Gamma$ are left free to vary independently in the different flux intervals (hereafter Model A). 

%
\begin{table}
\caption{Best-fitting Model A results ($\chi^2_r$/dof = 2.5/218)}           
\label{table:1}    
\begin{center}                        
\begin{tabular}{c c c c}      
\hline\hline               
Spectrum & T$_{BB}$             & $\Gamma$              & Norm.\tablefootmark{a} \\  
                & (eV)                  &                               &                                                 \\
\hline                      
   HF           & $2.00 \pm 0.06$       & $2.66 \pm 0.01$       & $18.8 \pm 0.6$                    \\    
   MF   & L-HF\tablefootmark{b}         & $2.86 \pm 0.02$       & $7.3 \pm 0.4$                     \\
   LF   & L-HF\tablefootmark{b}                 & $3.01 \pm 0.01$       & $3.4 \pm 0.1$                           \\
\hline                                 
\end{tabular} \\
\end{center}
\tablefoot{
\tablefoottext{a}{in 10$^{-4}$ ph s$^{-1}$ cm$^{-2}$ keV$^{-1}$}
\tablefoottext{b}{Parameter value linked to the value for the HF spectrum}
}
\end{table}

   \begin{figure}
   \centering
\resizebox{\hsize}{!}{\includegraphics{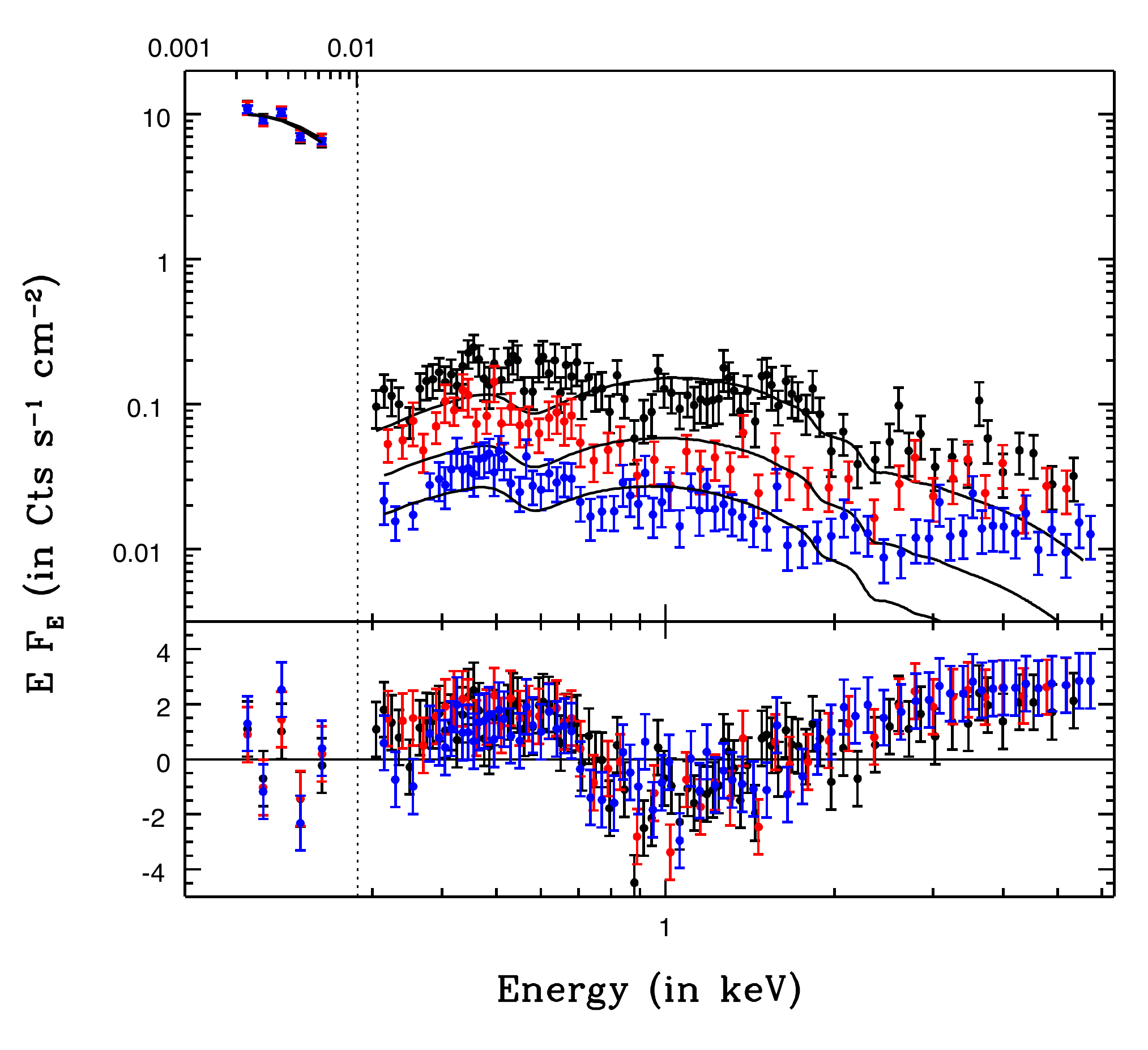}}
      \caption{Broad-band spectra of \pg\ in its HF (black), MF (red), and LF (blue) flux;  super-imposed are the corresponding 
best-fitting model A lines (top panel) and the residuals (in standard deviations) between the data and their best-fit models (bottom panel).}
         \label{Fig3}
   \end{figure}
   
Table 2 summarizes the best-fitting parameter values and statistics, for Model A. Figure 3 shows the HF, MF, and LF  spectra (black, red, and blue points, respectively).  The linesabove each spectrum indicate the best-fit Models A (top panel).  The plot in the bottom panel shows the residuals (in standard deviations) between the data and their best-fit models.  The model fits  the opt/UV part of the spectra well; however, it fails to accurately model the X--ray data, yielding  a global reduced $\chi^2$ ($\chi^2_r$)  of 2.5 for 218 degrees of freedom (dof).  For all  three flux intervals, the residuals in the X-rays show clear systematic deviations; there are two pronounced excesses between 0.3-0.7 keV and 2-6 keV, and a deficit in between (Fig.\,3, bottom panel). This is a clear signature, in low-resolution CCD-like spectra, of the presence of an ionized absorber, attenuating the primary X-ray continuum mainly at energies around 1 keV where the residual opacity due to highly ionized metals is stronger. This is not a surprising result, given the numerous previous  reports for the presence of such  material in \pg.


\subsection{Adding the warm-absorber}

%
\begin{table*}
\caption{Best-fitting Models B and C: $\chi^2_r/dof = 0.9/208$ and $\chi^2_r/dof = 0.9/212$, respectively}           
\label{table:1}    
\begin{center}                       
\begin{tabular}{c  c c c c c c c c}      
\hline             
\multicolumn{8}{c}{Model B} \\
\hline
Spec.   & $\Gamma$      & Norm.\tablefootmark{a}         & log$U$       & logN$_H$        & $f_c$         & $\upsilon_{out}$      
& F$_{Ion}\tablefootmark{b}     $                                               & F$_{2-10}$\tablefootmark{b} \\  
                &                        &                      &                       & (cm$^{-2}$)     &               & (km s$^{-1}$) 
&       &  \\ 

\hline                      
   HF  & $2.47 \pm 0.01$ & $4.4 \pm 0.1$ & $1.51_{-0.01}^{+0.02}$ & $23.21_{-0.02}^{+0.01}$ & $0.84 \pm 0.02$ & 
$-7100_{-8500}^{+5300}$ & $1.04 \pm 0.02$ & $0.057 \pm 0.001$ \\ 
   MF & $2.53 \pm 0.01$ & $3.1 \pm 0.1$ & $1.400_{-0.002}^{+0.033}$ & $23.20 \pm 0.01$& $0.88_{-0.03}^{+0.01}$ & L-HF & 
$0.86 \pm 0.03$ & $0.036 \pm 0.001$ \\
   LF  & $2.63 \pm 0.01$ & $1.9 \pm 0.1$ & $1.306_{-0.004}^{+0.032}$ & $23.20_{-0.02}^{+0.01}$ & $0.87\pm0.01$ 
& L-HF & $0.68 \pm 0.02$ & $0.019 \pm 0.001$ \\
\hline                                 
\multicolumn{8}{c}{Model C} \\
\hline
Spec. & $\Gamma$        & Norm. $^{1}$                                                                                  & log$U$ & logN$_H$       & $f_c$         & $v_{out}$ 
& F$_{Ion}^{2}$                                         & F$_{2-10}^{2}$ \\  
                &                       &       &               & (cm$^{-2}$)         &               & (km s$^{-1}$)
&       &  \\ 
\hline                      
   HF           & $2.47 \pm 0.01$ & $4.4 \pm 0.1$ & $1.51_{-0.01}^{+0.02}$ & $23.20_{-0.02}^{+0.01}$ & $0.87 \pm 0.02$ & 
$-3000$ (f) & $1.04 \pm 0.02$ & $0.057 \pm 0.001$ \\ 
   MF & $2.54 \pm 0.01$ & $3.0 \pm 0.1$ & $1.404_{-0.002}^{+0.033}$ & L-HF$^{3}$ & L-HF & L-HF & 
$0.86 \pm 0.03$ & $0.035 \pm 0.001$ \\
   LF  & $2.63 \pm 0.01$ & $1.8 \pm 0.1$ & $1.317_{-0.004}^{+0.032}$ & L-HF & L-HF & L-HF & 
$0.67 \pm 0.02$ & $0.019 \pm 0.001$ \\
\hline                                 
\end{tabular}
\end{center}
\tablefoot{
\tablefoottext{a}{in 10$^{-3}$ ph s$^{-1}$ cm$^{-2}$ keV$^{-1}$ at 1 keV};
\tablefoottext{b}{in $10^{-10}$ erg s$^{-1}$ cm$^{-2}$}
}
\end{table*}

We therefore added a dust-free ionized
absorber component (i.e. that does not affect the opt/UV data) to our Model A by multiplying the 
continua of the three flux-state spectra by three distinct {\tt phase} components (our PHoto-ionized Absorber Spectral Engine model, Krongold et al. 2003, implemented as a user-model in {\em Sherpa}), one for each flux interval (hereafter Model B). 
The parameters of a {\tt phase} component are (a) the ionization parameter $U$, defined as the ratio between the 
volume density of ionizing photons (i.e. with energy $\ge 13.6$ eV)   at the illuminated face of the ionized cloud of gas, $n_{phot}$, over the electron 
volume density in the cloud, $n_e$: $U = [Q_{ion}/(4 \pi c R^2)] / n_e = n_{phot} / n_e$ (where $Q_{ion}$ is the total number of ionizing photons); (b) the equivalent H column density of the cloud along our line of sight, N$_H$; (c) the inflow/outflow radial velocity 
of the absorber along the line of sight, $\upsilon_r$; and (d) the fraction of ionizing source covered by the absorber along the direction of the observer, $f_c$. 

We first left all  three of the relevant parameters, $U$, N$_H$, 
and $f_c$, free to vary independently between the three {\tt phase} components, while the radial velocity of the absorber was kept linked to a 
single value in all  three spectra and initially left free to vary in the fit.  As before with Model A, T$_{BB}$ was forced to be the same for all  three flux-state spectra, while $N$ and $\Gamma$ were left free to vary independently in the different flux-state spectra. 

Table 3 summarizes the best-fit parameter values and statistics for Model B, which provides an excellent description 
of both the UV and X-ray data in all  three flux intervals.  The best-fit soft-photon temperature is T$_{BB}=1.76 \pm 0.03$ eV. In columns 8 and 9 we list the model ionizing flux (i.e. the unabsorbed 0.0136-100 keV continuum flux) and the model unabsorbed X--ray flux in the 2--10 keV band. 

The model results indicate that neither the column density nor the covering factor varies (within the errors) between the three flux-state spectra.  Moreover, 
while the sign of the line of sight velocity of the absorber indicates that the gas is outflowing (even at our low resolution we rule out positive velocities 
at $\simeq 95$\% confidence level; see Fig. 4), our low-resolution X-ray spectra cannot constrain this velocity to better than $\sim 70$\% (see Fig.\,4). 
Our best-fit velocity is consistent with the accurate value of $\upsilon_r = -(3000 \pm 500)$ km s$^{-1}$ 
reported by Kaspi \& Behar (2006), based on the high-resolution XMM-{\em Newton}-RGS spectrum of PG~1211+143, but only marginally consistent 
(within the $3\sigma$ level) with the $v_r \sim -22000$ km s$^{-1}$ value reported by Pounds (2014) for the mildly relativistic outflow in this object 
(see Fig.\,4). Our  0.3-6 keV spectra are not sensitive to the even higher velocity $v_r \simeq 0.14c$ outflow first reported by Pounds \& Page (2006). 

   \begin{figure}
   \centering
\resizebox{\hsize}{!}{\includegraphics{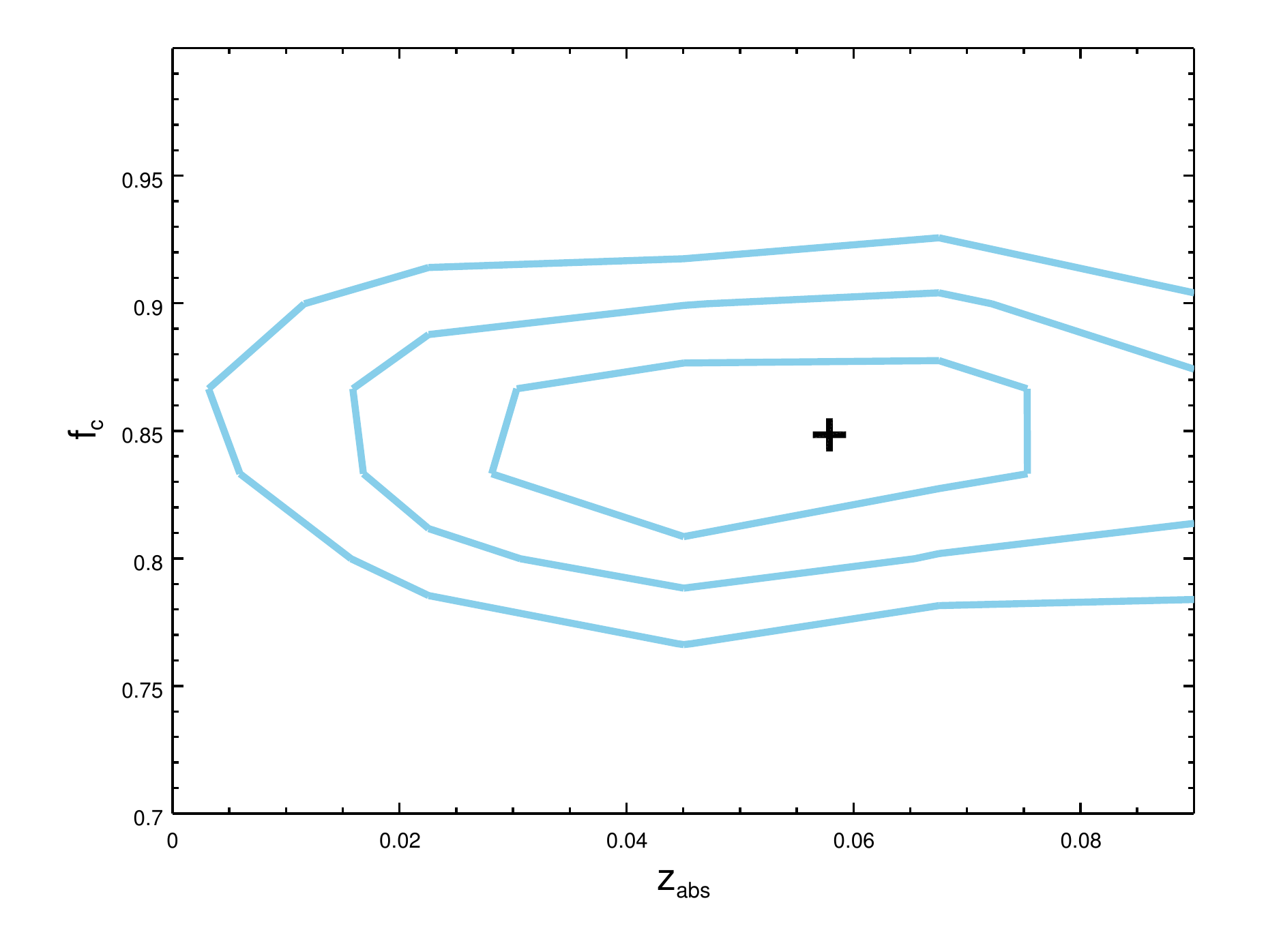}}
      \caption{Contours at 1, 2, and 3$\sigma$ confidence levels  for the best-fitting redshift of the ionized absorber of Model B, which translates into 
the radial velocity of the absorber through the relationship $\upsilon_r = (z_{abs} - z_{sys})$ * c, where $z_{abs}$ and $z_{sys}$ are respectively the redshift of  the absorber and the systemic redshift of \pg, and c is the speed of light.}
         \label{fc_vs_z}
   \end{figure}
   
We therefore linked the column densities and covering factors of the absorber to a single value in all three flux-state spectra, we froze the outflow velocity of the absorber to $v_r = -3000$ km s$^{-1}$  (Model C), and we repeated the fit. The best-fitting parameters and statistics of Model C are given in Table 3, while Fig.\,5 shows the spectra with the corresponding best-fitting models (top panel) and residuals (bottom panel).  Model C fits the UV and X-ray data in all  three flux intervals as accurately as  Model B ($\chi^2_r/dof = 0.9/212$). The best-fit soft-photon temperature is again T$_{BB}=1.76 \pm 0.03$ eV. The residuals are flat over the entire energy range covered by our data (Fig.\,5, bottom panel). 

   \begin{figure}
   \centering
\resizebox{\hsize}{!}{\includegraphics{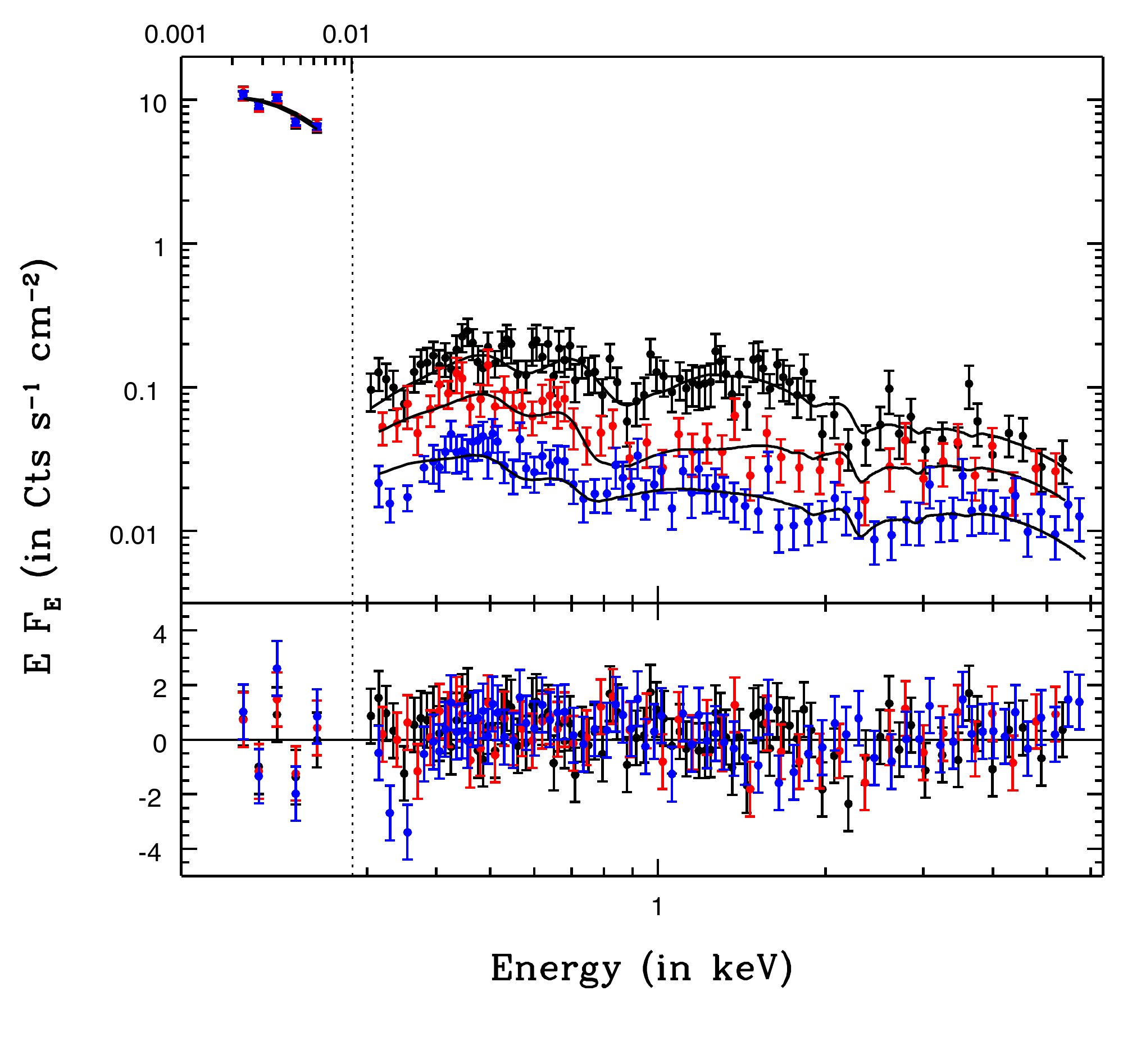}}
      \caption{Same as Fig.\,3, for the Model C best-fit results.}
         \label{Fig5}
   \end{figure}
   
We note that both the ionization parameter of the photo-ionized outflow and F$_{Ion}$ decrease monotonically from the HF to the LF spectrum. 
Figure 6 shows the measured number of ionizing photons per steradian emitted by the central source, $Q_{ion}$ (divided by $4\pi c$), plotted against the best-fit ionization parameter $U$, for the three flux-interval spectra, in log-log space. Clearly the two quantities are linearly correlated (as shown by the best-fit line in the same figure). The ionized absorber is thus consistent with being always in photo-ionization equilibrium with the ionizing flux during the entire {\em Swift} campaign. This allows us to infer an accurate estimate on the quantity $(n_eR^2)$. Using the definition of $U$, it can be shown that log$(Q_{ion}/4\pi c$) = log$U$ + log$(n_e R^2)$. Therefore, the intercept of the best-fit line shown in Fig.\,6 is a measure of the product between the electron volume density in the absorbing gas and the square of the distance of the cloud from the ionizing source. In this way, we find that  $n_e R^2 = 2 \times 10^{42}$ cm$^{-1}$.

\begin{figure}
   \centering
\resizebox{\hsize}{!}{\includegraphics{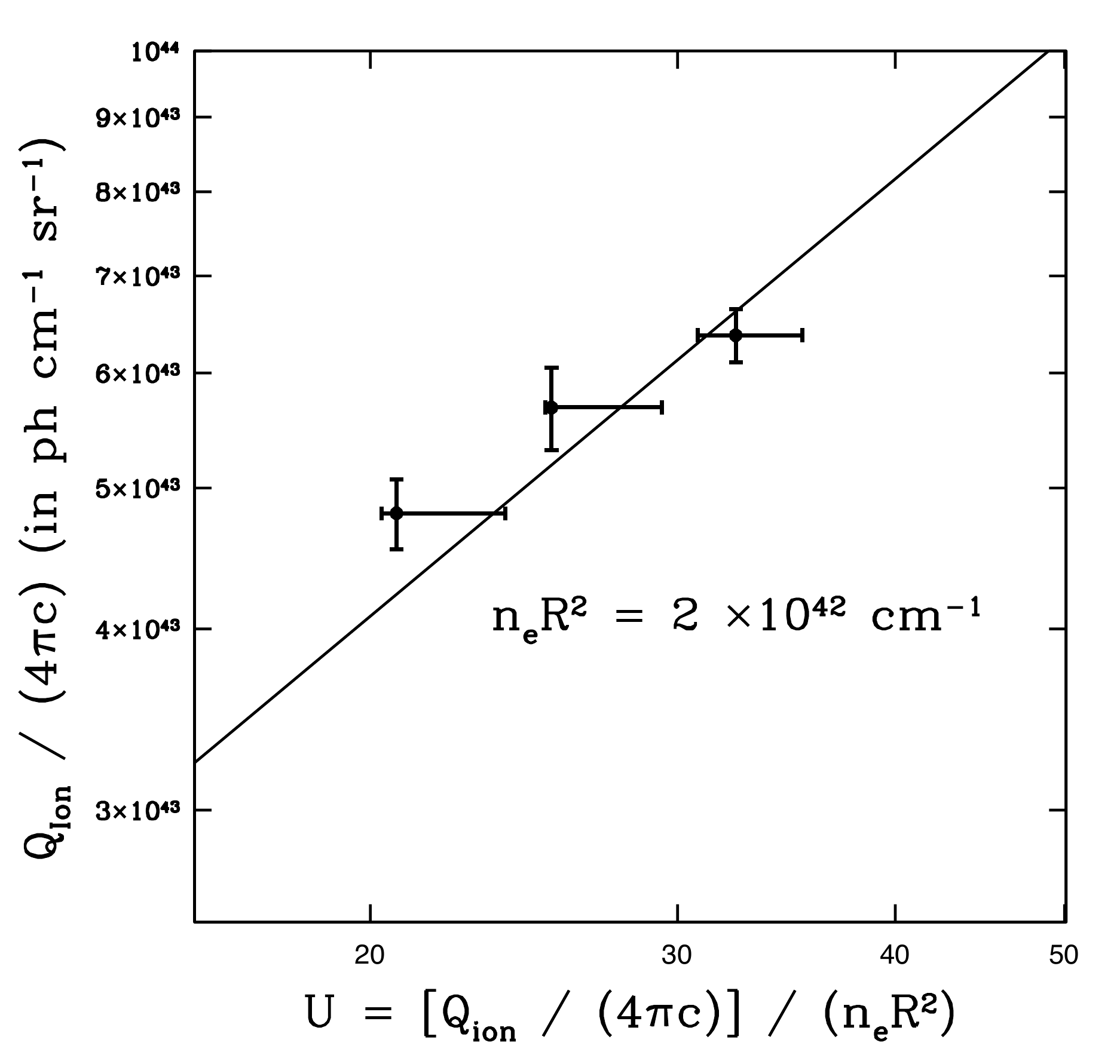}}
      \caption{Best-fitting number $Q_{ion}$ of ionizing photons at the illuminated face of the absorbing cloud of gas (divided by the quantity $4 \pi c$), vs 
the best-fitting ionization parameter, $U$, for the three flux-interval spectra HS, MS and LS. Both the $x$ and $y-$axis are in logarithmic scale. 
The straight line indicates  the best-fitting linear relationship between log($Q_{ion}$) and log($U$), whose intercept provides an estimate for the quantity $n_e R^2)$ (see \S4.2).}
         \label{ner2}
\end{figure}
   
\subsection{Density, distance, and structure of the outflow}
The photo-ionization equilibrium scenario also allows us to remove the degeneracy between $n_e$ and $R$, by setting a lower limit on $n_e$. If the \pg\ absorber is in photo-ionization  equilibrium during the entire campaign, the absorbing gas must be able to relax to photo-ionization  equilibrium over a time scale shorter than the shortest time interval between two consecutive different X-ray flux intervals of the quasar.

The time needed for a cloud of gas to reach equilibrium with its ionizing radiation depends inversely on the gas volume density, both during increasing and decreasing phases of the ionizing radiation. It is typically significantly longer during recombination than during ionization phases 
(e.g. Nicastro et al. 1999; Krongold et al. 2007). For a 3-dominant-ion condition (i.e. when the ion fractions of a given element are mostly distributed among 3 consecutive species) this equilibration time can be analytically approximated as   
$$t_{eq}^{X^i,X^{i+1}}(t\to t+dt) \sim$$ 
$${\left[{1 \over 
{\alpha_{rec}(X^i,T_e)_{eq} n_e}}\right]}  
{\left[{1 \over {\left({{\alpha_{rec}(X^{i-1},T_e)}  \over 
{\alpha_{rec}(X^i,T_e)}}\right)_{eq}  + 
\left({n_{X^{i+1}} \over {n_{X^i}}} \right)_{eq}}}\right]_{t+dt}}, 
$$
where $t_{eq}^{X^i,X^{i+1}}$ is the equilibration time scale relative to the transition from the ions $i$ and $(i+1)$ of the element $X$, as the source flux changes over the time interval from $t$ to $(t+dt)$; $\alpha_{rec}(X^i,T_e)$, $n_e$, and $n_{X^i}$ are the radiative-recombination coefficient of the ion $X^i$ at an electron temperature $T_e$, the gas electron volume density, and the relative fraction of the ion $X^i$, respectively. The subscript $eq$ indicates quantities at equilibrium at the time $(t+dt)$.

In our case, the 3-dominant-ion hypothesis is verified for several light elements. For example, our Model C best-fit results show that the three highest ionization ions of  oxygen (He-like, H-like, and fully-stripped) have three-digit approximated fractions of $f_{OVII} = 0.277$, $f_{OVIII} = 0.325$, and $f_{OIX} = 0.398$ (log$U = 1.317$: LS); $f_{OVII} = 0.042$, $f_{OVIII} = 0.287$, and $f_{OIX} = 0.671$ (log$U = 1.404$: MS); and $f_{OVII} = 0.002$, $f_{OVIII} = 0.106$, and $f_{OIX} = 0.892$ (log$U = 1.51$: HS), and that their sum is equal to unity in all cases. During the {\em Swift} campaign, the shortest time interval for a transition from  one flux-interval to a another is that between the fourth (flux-interval MS) and  fifth (flux-interval LS) observations, separated by $\Delta t = 1$ day (86400 s). 

We can then evaluate the quantities in equation (1) for the three highest ionization ions of oxygen at the end of the transition, i.e. in the flux-state LS. The temperature of photo-ionized gas with log$U$(LS)$= 1.317$ (and logN$_H = 23.2$, in cm$^{-2}$) is $T$(LS)$= 2.5 \times 10^5$ K. This gives $\alpha_{rec}(OVIII;2.5\times 10^5) = 2.5 \times 10^{-12}$ cm$^3$ s$^{-1}$, $\alpha_{rec}(OVII;2.5\times 10^5) = 3.8 \times 10^{-12}$ cm$^3$ s$^{-1}$ (coefficients from Shull \& van Steenberg, 1982), and $(n_{OIX}/n_{OVIII}) = 1.22$. Thus, 
\begin{equation}
t_{eq}^{OVIII,OIX}(4^{th} \to 5^{th} observation) \sim 1.5 \times 10^{11} n_e^{-1} < 86400 s 
\end{equation}
\noindent
or $n_e > 1.7 \times 10^6$ cm$^{-3}$. From $(n_eR^2) = 2 \times 10^{42}$ cm$^{-1}$, we then get $R < 1.1 \times 10^{18}$ cm $=0.35$ pc (or 1.1 lt-years). This radius is fully consistent with the broad line region (BLR) radius in this object, which is of the order of $\sim 0.3$ lt-years (Kaspi et al. 2000).

Following Krongold et al. (2007), and assuming homogeneity in the flow, we can then estimate the line-of-sight thickness of the outflow, $\Delta R$. This is simply equal to the ratio of the equivalent H  column density over the H volume density, $n_H$.  For a fully ionized gas (which is a good approximation here), $n_H$ is equal to $n_e/1.23$. Therefore, using the best-fit $N_H$ estimate and the upper limit on $n_e$ that we estimated above, we find that $\Delta R = N_H/n_H \simeq 1.23 N_H / n_e < 1.1 \times 10^{17}$ cm $=0.035$ pc (0.11 lt-years). We can also provide an upper limit on the relative thickness of the out flow as follows: $(\Delta R / R) = (N_H / n_H) / R = 1.23 N_H (n_eR^2)^{-1/2} (n_e)^{-1/2}$. Using our $n_eR^2$ estimate (see Section 4.3), our best-fit $N_H$ value, and the lower limit on $n_e$, we find that $\Delta R/R < 0.1$. 

\section{Discussion and conclusions}

We present the results of a detailed study of the data from the  {\it Swift}  monitoring campaign of \pg\ in 2007. This is a bright quasar, with a reverberation BH mass estimate of $\sim 1.5\times10^8$ M$_{\odot}$ (Peterson et al. 2004). Based on our best-fit Model C results, the $1\mu m-100$ keV luminosity of \pg\  (which effectively measures the bolometric luminosity, $L_{bol}$) is of the order  of $5 \times 10^{45}$ erg s$^{-1}$. This is $\sim 0.4$ of the Eddington limit for a $10^8$ solar mass BH. 

We studied the broad-band spectra of the source by constructing opt/UV/X--ray  spectra in three different X--ray flux intervals. Our main aim was to describe the broad-band SED with the simplest  physical model possible, and to use the observationally strong fact of highly variable X--ray emission in the  absence  of opt/UV variations to constrain the model as much as possible. We found that the three SEDs are modelled accurately by two physically motivated, spectral components. The first  accounts for the effects of the warm absorber in this object. The second  accounts for the opt/UV/X--ray emission. We find that a single {\tt nthcomp} model, with a constant temperature of the input soft photons, can fit  the opt/UV/X-ray SED well in all three X--ray flux levels. From a physical point of view, this result corresponds to the emission from an accretion disc consisting of multiple black-body components and Comptonization from a hot corona which covers the disc. 

The configuration of the inner region in \pg\ could be similar  but not identical to what has been observed in other AGN as well. For example, Petrucci et al. (2013) presented the results from the study of the broad-band SED of Mrk 509. The physical picture in \pg\ could be similar to the picture shown in their  Fig.\,10. The deep layers of the accretion disc radiate like a black body at $T_{BB}\sim 2$ eV, and there is a hot layer on top of the disc  (we cannot constrain its temperature with the current data; we assumed a temperature of 100 keV), which upscatters the optical-UV photons to form the X--ray spectrm. Basically, the upper disc layer in \pg\ appears to be much hotter than the {\tt WARM} corona in Mrk 509 (as Petrucci et al. called it), and could be the {\tt HOT} corona itself. 

A soft-excess spectral component is not necessary to fit the present data set. In fact, evidence that this component is missing  in \pg\ is quite strong and is independent of our spectral model fit assumptions. This evidence results from the fact that the soft and the hard band count rates vary proportionally most of the time, and the soft-vs-hard counts plot is well fitted by a straight line with an intercept consistent with zero. Noda et al. (2011, 2013)  have detected significant positive offsets in similar soft-vs-hard ``count-coubt'' plots in a handful of X--ray quasars. They have interpreted them as evidence  of an extra soft-excess component which is less variable than the X--ray continuum. The lack of a similar non-positive offset in our  case argues against the presence of an extra, less-variable, soft-component in this source. 

Our results strongly suggest that both the soft and hard band X--ray photons are produced by the same component, hence the linear relation between the two. At the highest X--ray flux intervals the lineal relation breaks. The soft X--ray count rates increase by a factor that is larger than in the hard band. As we demonstrate and discuss in \S 5.3, this is most probably due to strong opacity changes, caused by higher ionization of the warm absorber, which affect mainly the 0.3--2 keV energy band, and not the X--ray photons at higher energies.

\subsection{ X--ray and opt/UV variability}

We do not detect any significant optical/UV variability during the 80-day  observations. The lack of opt/UV variations is consistent with the high BH mass measured in this object, as the various disc time scales are expected to be long in this case. For example, the viscous time scale for an accretion disc around a BH is given by the relation $t_{visc}\sim 10^5\alpha_{0.1}^{-1}R_{3}^{3/2}M_8(r/h_d)^2$ s, where $\alpha_{0.1}$ is the viscosity parameter (in units of 0.1), $R_3$ is the disc radius (in units of 3 Schwarzschild radii, $R_{S}$), $M_8$ is the BH mass (in units of $10^8$ M$_{\odot}$), and $r/h_d$ is the ratio of radius over the disc thickness (Czerny, 2006). For a thin disc (i.e. a disc where $h_d/r\sim 0.1$), the viscosity time scale is of the order of $\sim 120$ days for $M_8\sim 1$ at a radius equal to 3$R_S$ (i.e. the innermost radius for a stable circular orbit in a non-rotating BH) assuming a value of 0.1 for the viscosity parameter. This time scale is longer than the duration of the \Sw\ monitoring campaign of \pg.

On the other hand, the soft and hard X--ray band count rates are highly variable at all time scales. We believe that the simultaneous absence of significant opt/UV variations is not consistent with the hypothesis of inwards propagating disc accretion rate fluctuations. The absence of opt/UV variations argues against the operation of accretion rate fluctuations at larger disc radii. Therefore, the long term X--ray variations cannot be due to propagating fluctuations that happened at larger radii. Of course, the propagation time scale may be longer than the time span of the 2007 \Sw\ observations, and the X--ray variations we observe may be due to fluctuations at larger radii which happened before the start of the observations.  It is possible that we were somewhat unlucky and that disc fluctuations at large radii simply did not occur during the 2007 observations, although we consider it rather unlikely.

Furthermore, X--ray illumination does not seem to  significantly affect the disc emission in \pg\ (as again, we would expect the opt/UV flux to be variable). This is not surprising, given the small fraction of the X--ray over the bolometric luminosity (which is mainly determined by the disc flux). The ratio between the $F_{ion}$ and $F_{2-10}$ fluxes (listed in Table 3) is of the order of 18--36 in the HF and LF intervals (given the steep X--ray spectrum, the $F_{2-10}$ flux is representative of the total X--ray flux in this object). Therefore, energetically, the X--rays are not powerful enough to influence the opt/UV emission of \pg. 

\subsection{Constraints on the disc and X--ray corona}

According to our results the temperature of the inner disc radius, T$_{BB}$, remains constant and equal to $\sim 2$ eV$\sim 2.3\times 10^4$K. Assuming that the opt/UV emission in \pg\ is produced
by a multicolour accretion disc, the hottest temperature in the disc should be $\sim 0.5(3GM\dot{M}/8\pi R_{\rm in}^3\sigma)^{1/4}$ (Pringle, 1981). Assuming a BH mass of $\sim 10^8$ M$_{\odot}$, $R_{\rm in}=3R_S$, an accretion rate of $\sim 0.4 \dot{M}_{\rm Edd}$, and an efficiency of 0.1, we find that the maximum disc temperature should be of the order of $\sim 9.5$ eV, which is larger than the soft photon temperature of $\sim 2$ eV that we find. This temperature, for the same physical parameters, corresponds to a disc radius of $\sim 25R_S$. 

The accretion flow at smaller radii may be filled by the hot X--ray corona. The X--ray spectral slope we found ($\sim 2.5-2.6$) is rather steep, but is in agreement with the measurement of Zoghbi et al. (2015). We cannot constrain the temperature of the X--ray corona in \pg,  but we do observe normalization and spectral slope variations.  The spectral shape of Comptonization models depends on the temperature and optical depth of the corona. We detect a $\Delta \Gamma \sim 0.15$ steepening of the spectral slope in the 2--10 keV band, as the flux decreases by a factor of 3, from the high to the low flux state. Interestingly, this ``steeper when fainter'' is uncommon in Seyferts, where we usually observer a ``steeper when brighter'' behaviour  (i.e. Sobolewska \& Papadakis, 2009). This is further evidence for the decoupling between the opt/UV and the X--ray emission. A variable opt/UV soft photon input would result in spectral slope -- flux relation opposite to what we observe.
The {\tt nthcomp} model  does not provide an estimate for the optical depth of the hot plasma. We compared the {\tt nthcomp}  best-fit spectral models with those from the {\tt compps} model (Poutanen \& Svensson, 1996), assuming a slab geometry and the same T$_e$ and T$_{BB}$ values. We found that the observed spectral slopes correspond to an optical depth of $\tau \sim 0.15-0.1$ (for the HF and LF spectra, respectively). 

We note that {\tt nthcomp} does not include an explicit factor for the relative power between the disc and the corona emission. We can get a rough estimate of the ratio of the energy released in each phase following Appendix B of Petrucci et al. (2013). Using  their equation (B.4), we find that the soft photon count rate, $n_s$, is roughly equal to the observed photon count rate (since $\tau$ is much smaller than unity). The observed count rate can by computed by the best-fit {\tt nthcomp} model (we assume here an average $\Gamma$ of 2.5, an average Norm of $3\times 10^{-3}$, and $kT_e=100$ keV). Knowing $n_s$ and the temperature of the disc modified black-body spectrum, we can estimate the soft photon luminosity, $L_s$. Using equation (B10) of Petrucci et al. (2013), we can then estimate the heating power, $L_h$, liberated in the corona to Comptonize  the soft photon field. In this way, we found that the amplification ratio, $A=L_h/L_s$, is roughly equal to 2. This is the case when all the gravitational power is liberated in the hot medium (Haardt \& Maraschi, 1991). However, if this were the case, we would expect the disc emission to be entirely due to X--ray reprocessing, and hence, to be variable as well, which is not the case. If the X--ray emission region is outflowing, so that the photons Comptonized in the corona are only emitted upward, then  $L_h\approx L_{obs}-L_s$ and $A\approx1$, which is the case when the gravitational power is released almost entirely in the disc. Consequently, our results suggest the case of an outflowing corona in \pg.

\subsection{ Warm-absorber and  X-ray emission}

The X--ray fractional variability amplitude is of the order of $\sim 0.6-0.85$, which is quite large. For comparison, the maximum observed fractional variability amplitude of nearby bright Seyferts is of the order of $\sim 0.4-0.5$ over time scales of a year or so (Zhang 2011). This is smaller than the value we observe in \pg\ over a period of just $\sim$2 months. This enhanced variability amplitude is probably due to the warm absorber variations. Indeed, our best-fit Model C results predict a 2--10 keV intrinsic flux variation by a factor of $\sim 3$. This is not enough to explain the observed max-to-min variability by a factor of $\sim 5$ and $10$ in the hard and soft band, respectively. This occurs because the X--ray central source is further absorbed, in situ, by a dust-free ionized outflow crossing our line of sight to the quasar and dominating the observed variations in the soft band. 

According to our results, the absorber covers almost 90\% of the central source. It is outflowing with a velocity of less than 2.3$\times10^4$ km/s (upper 3$\sigma$ limit) and has a column density of $\log N_{H}\sim 23.2$. It is in photo-ionizing equilibrium with the ionizing flux and  is located at a distance of less than 0.35 pc from the central source. The relative thickness of the absorber, $\Delta R/R$, is less than 0.1. Therefore, the \pg\ outflow, as seen  along our line of sight is significantly thinner than its distance from the central ionizing source. This suggests, at least at the location where our line of sight crosses the flow, that the outflow is not propagating outwards radially  but rather transversally with respect to our line-of-sight perspective.

   \begin{figure}
   \centering
\resizebox{\hsize}{!}{\includegraphics{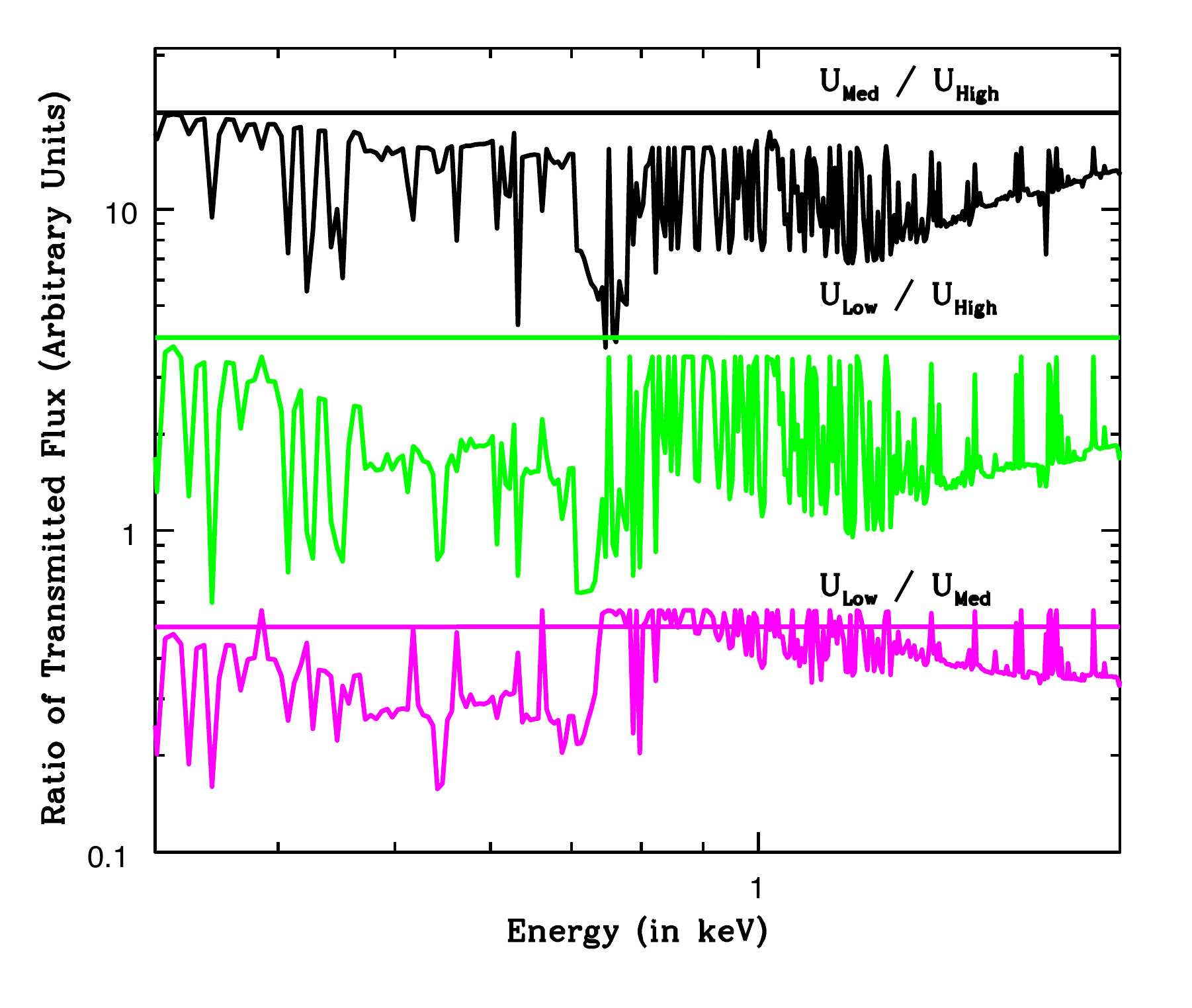}}
      \caption{Ratios between the best-fitting transmitted models (at high spectral resolution of $\Delta E = 5$ eV) of the MF/HF (black), LF/HF (green), and LF/MF (magenta) spectra in the 0.3-2 keV spectral range (see text for details).}
         \label{transm}
   \end{figure}

The opacity of an ionized absorber to X-ray photons depends critically on the degree of ionization of the absorber itself.  During the {\em Swift} campaign, the ionized absorber varies between log$U\simeq 1.3-1.5$. We expect both the He- and H-like ions of the elements lighter than S, as well as the Fe unresolved transition array (UTA) and the higher ionization Fe L transitions, to play an important role in modifying the transmitted spectral shape of the continuum between E$\simeq 0.3-3$ keV. This is clearly shown in Fig.\,7 where we plot the ratios between the best-fitting transmitted models (at high spectral resolution of $\Delta E = 5$ eV) of the spectra MF/HF (black), LF/HF (green), and LF/MF (magenta) in the 0.3-2 keV spectral range. The straight lines are the identity ratios for the respective cases (the three ratios have been multiplied by factors of 200 (black), 40 (green), and 0.5 (magenta) to separate them and make them clearly visible in a single panel.)

The changes in opacity between the three flux-state affect the entire 0.3-2 keV spectral range, and clearly extend even at energies higher than 2 keV. The strongest variations are between the HF and the LF spectra (green curves), where the high-ionization Fe L opacity changes mimic  a flattening of the spectral slope at E$\gtrsim 1.2$ keV (at the low resolution
of our CCD spectra), while the dramatic changes in the relative fraction of iron UTA ions and in the C-Ne K region result in the deepening of the trough between 0.4-0.8 keV during the LF. The latter is clearly also visible in the transition between the MF and the LF (magenta curves), while a smoother change in the overall spectral shape between 0.5 and $\gtrsim 2$ keV is induced by the change of opacity of the absorber between the HF and the MF intervals (black curves). These opacity changes are most probably responsible for the 
deviation from the linear soft vs hard X-ray count rate relation at high X-ray fluxes (Fig. 2).

\subsection{Implications for feed-back} 

Based on our best-fitting Model C estimates of the physical (log(N$_H) = 23.20$),  kinematic ($v_r = -3000$ km s$^{-1}$), and geometrical 
(thin flow with R$< 0.35$ pc and $\Delta R / R < 0.1$) parameters of the  ionized outflow of PG 1211+143,  we can estimate the mass outflow 
rate of this moderate-velocity outflow. 
We assume the same conical disc-wind geometry that  Krongold et al. (2007) adopted for the low-velocity  outflow of the low-luminosity Seyfert 1 galaxy NGC~4051, where  all the evidence also pointed to a transverse geometrically thin absorber. Following Krongold et al. (2007), the mass outflow rate for this geometry can be estimated as $\dot{M}_{out} \simeq 1.2 \pi m_p N_H v_r R < 3.2 \times 10^{26}$ g s$^{-1} \simeq 5$ M$_{\odot}$ yr$^{-1}$ (assuming a vertical disc wind and an inclination angle of 30 degrees). This upper limit is about 2.3 times larger than the Eddington accretion rate for \pg\ (which is about 2.2 M$_{\odot}$ yr$^{-1}$, assuming $L_{bol}=5\times 10^{45}$ ergs/s, and a 10\% radiative efficiency). If the mass outflow rate is indeed that high, then outflows like this are probably short-lived episodes in the quasar lifetime. 

Finally, for the kinetic power of the outflow we obtain $P_{out} = 1/2 \dot{M}_{out} v_r^2 < 1.4 \times 10^{43}$ erg s$^{-1}$, i.e. only about 7\% of the 
radiative power of the quasar. At this power, it would require about 2.3 Gyrs for the outflow to deploy enough mechanical energy ($\simeq 10^{60}$ ergs) into the galaxy and the surrounding inter-galactic medium (IGM)  to control their evolution. This is probably far too long for a plausible wind lifetime. However, it would only require about 3 Myrs for such an outflow to deposit enough energy (about $10^{57}$ ergs, assuming an average ISM density of $n_H = 0.01$ cm$^{-3}$; e.g. Krongold et al. 2007 and references therein)  in the 100 kpc radius, 1 kpc thick disc of its host galaxy, and heat the ISM to its evaporation temperature of $\simeq 10^7$ K. This is probably sufficient to produce a fast decline in the star formation rate of the galaxy. 
 
\begin{acknowledgements}
This work was supported by the ``AGNQUEST'' project, which is implemented under the "Aristeia II" Action of the "Education and lifelong Learning" operational programme of the GSRT, Greece.
\end{acknowledgements}

\bibliographystyle{aa}

\end{document}